\documentclass[]{spie}

\usepackage{amsmath,amsfonts,amssymb}
\usepackage{graphicx}
\usepackage[colorlinks=true, allcolors=blue]{hyperref}
\usepackage{xfrac}

\usepackage{tikz}
\usetikzlibrary{arrows,shadows,positioning}

\tikzset{
  frame/.style={
    rectangle, draw,
    text width=6em, text centered,
    minimum height=3em,drop shadow,fill=white,
    rounded corners,
  },
  line/.style={
    draw, -latex',
  },
  mynode/.style={
    draw, left=2pt, align=center,
  }
}

\newcommand{\myfraction}{0.75}

\title{A Deep Reinforcement Learning Approach to Wavefront Control for Exoplanet Imaging} 

\author[a,b,c]{Yann Gutierrez}
\author[a]{Johan Mazoyer}
\author[b]{Olivier Herscovici-Schiller}
\author[c]{Laurent M. Mugnier}
\author[b]{Baptiste Abeloos}
\author[a, d]{Iva Laginja}
\affil[a]{LESIA, Observatoire de Paris, Université PSL, Sorbonne Université, Université Paris Cité, CNRS, 5 place Jules Janssen, 92195 Meudon, France}
\affil[b]{DTIS, ONERA, Université Paris Saclay, 91120 Palaiseau, France}
\affil[c]{DOTA, ONERA, Université Paris Saclay, 92320 Châtillon, France}
\affil[d]{NOVA/Leiden University, Einsteinweg 55, 2333 CC Leiden, The Netherlands}

\authorinfo{Further author information: (Send correspondence to Y.G.)\\Y.G.: E-mail: yann.gutierrez@obspm.fr, yann.gutierrez@onera.fr}

\begin{document}
\maketitle

\begin{abstract}
Exoplanet imaging uses coronagraphs to block out the bright light from a star, allowing astronomers to observe the much fainter light from planets orbiting the star. However, these instruments are heavily impacted by small wavefront aberrations and require the minimization of starlight residuals directly in the focal plane. State-of-the art wavefront control methods suffer from errors in the underlying physical models, and often require several iterations to minimize the intensity in the dark hole, limiting performance and reducing effective observation time. 
This study aims at developing a data-driven method to create a dark hole in post-coronagraphic images. 
For this purpose, we leverage the model-free capabilities of reinforcement learning to train an agent to learn a control strategy directly from phase diversity images acquired around the focal plane.
Initial findings demonstrate successful aberration correction in non-coronagraphic simulations and promising results for dark hole creation in post-coronagraphic scenarios.
These results highlight the potential of model-free reinforcement learning for dark-hole creation, justifying further investigation and eventually experimental validation on a dedicated testbed.

\end{abstract}

\section{Introduction}

    Space-based direct imaging of exoplanets enables the search for life signs through the characterization of their atmospheres and the estimation of their surface temperature. Because exoplanets are orders of magnitude fainter than their host star, coronagraphs are used to suppress starlight in order to improve the contrast. However, quasi-static aberrations, introduced by small imperfections in the optical components and slowly evolving thermo-mechanical distortions, cause some of the starlight to leak through the coronagraph. The starlight residuals create speckles in the scientific images, that are indistinguishable from potential planets, compromising detection capabilities.

    One approach to improve image quality involves the use of deformable mirrors (DMs) in conjunction with wavefront-sensing techniques. Focal-plane wavefront sensing approaches, such as Phase Diversity \cite{gonsalves1982} are particularly adapted for space telescopes lacking dedicated wavefront sensors, as they only require images taken by the scientific sensor to estimate the aberrations \cite{MUGNIER20061}. When coupled with a wavefront control algorithm, they can significantly reduce speckle intensity within a designated focal-plane region, known as the dark hole. State-of-the-art methods are currently limited by errors in the physical models they rely on, and require several iterations to minimize the intensity in the dark hole, reducing the time available for actual scientific observations (see section 5.4 of reference~\citeonline{GalicheretMazoyerreview}). 

    This work investigates the potential of model-free reinforcement learning (RL) as a fully data-driven approach for post-coronagraphic wavefront control. Unlike supervised learning approaches that focus only on estimating the aberrations \cite{quesnel_deep_2020}, our approach aims at controlling the DM direclty using images captured by the scientific sensor, specifically phase diversity images. Additionally, the normalized intensity within the dark hole serves as a natural optimization target for the RL algorithm, making it an interesting avenue for exploration.

    We started by developing a model-free RL method to perform non-coronagraphic wavefront correction directly in the focal plane, without additional hardware. While this task is inherently challenging, it presents a simpler scenario compared to post-coronagraphic control, as it is less sensitive to high-order aberrations, and amplitude aberrations can be neglected. The success achieved in this initial phase provided a strong foundation for transitioning to the more complex task of post-coronagraphic dark-hole creation, for which we present preliminary results. The method successfully learns to create a superficial dark hole, even without optimized training parameters. These findings highlight the promise of RL in this domain and motivate further exploration to refine the approach.

\section{Method}

    Here, we summarize the introduction to RL found in Gutierrez et al. (2024\cite{Gutierrez-a-24}) then formulate image-based wavefront control in the context of RL. Finally, we describe the algorithm used to train the agent.

    \subsection{Reinforcement learning}
    
        Reinforcement Learning \cite{sutton2018reinforcement} is a branch of machine learning focused on training intelligent agents to make decisions within dynamic environments with the goal of maximizing cumulative rewards. The agent interacts with an environment, receiving rewards or penalties based on its actions, aiming to learn a policy that optimizes long-term reward accumulation. This involves a balance between exploiting known actions for immediate rewards and exploring new actions to discover better strategies over time.
        
        A Markov Decision Process (MDP) provides the mathematical framework for RL, where an agent interacts episodically with an environment. Each episode consists of discrete time steps $t$ where the agent observes a state $s_t$ and selects an action $a_t$, which prompts the environment to emit a reward $r_{t+1}$ and transition to a new state $s_{t+1}$. The state-transition probability function $p$ determines the likelihood of moving from one state to another given an action. It follows the Markov property, meaning the next state only depends on the current state and the chosen action. In practice, agents often receive observations $o_t$ instead of the true state, providing indirect information about the environment's state.
        
        In order to make decisions effectively, the agent aims to optimize a policy $\pi$. This policy is a mapping from states to a probability distribution over actions, denoted as $\pi(a|s)$, which indicates the likelihood of selecting each action $a$ when in state $s$. By refining this policy over time through interactions with the environment, the agent seeks to improve its decision-making strategy, ultimately maximizing cumulative rewards in the MDP framework.
    
    \subsection{Image-based wavefront control as a reinforcement learning problem}
    
        In this subsection, we describe the wavefront correction problem in the context of RL, in the non-coronagraphic case. The changes enacted to accommodate for a coronagraphic system will be described in Section \ref{sec:coro}.
        
        In a non-coronagraphic setup, at wavelength $\lambda$, the image of a point source at infinity is degraded by the pupil plane phase $\phi$.
        
\begin{figure}[ht!]
            \centering
            \begin{tikzpicture}[font=\small\sffamily\bfseries,very thick,node distance = 4cm, every label/.style={align=center}]
                \node [frame] (agent) {Agent};
                \node [frame, below=2.65cm of agent] (environment) {Environment};
                \draw[line] (agent) -- ++ (3.5,0) |- (environment)  node[left, pos=0.115, align=right] {action\\ $a_t$}
                node[mynode,pos=0.41, anchor=south east] {\includegraphics[scale=0.16]{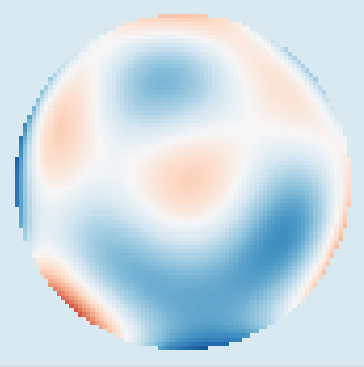}} node[left, pos=0.43] {DM command};
                \coordinate[left=8mm of environment] (P);
                \draw[thin,dashed] (P|-environment.north) -- (P|-environment.south);
                \draw[line] (environment.190) -- (P |- environment.190)
                node[midway,above]{$o_{t+1}$};
                \draw[line,thick] (environment.170) -- (P |- environment.170)
                node[midway,above]{$r_{t+1}$};
                \draw[line] (P |- environment.190) -- ++ (-1.5,0) |- (agent.170) node[left, pos=0.46, align=right] {observation \\ $o_t$}
                 node[mynode, pos=0, anchor=south east] {\includegraphics[scale=0.18]{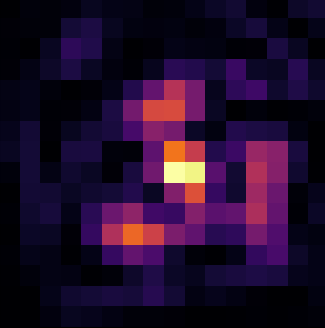} \\ \includegraphics[scale=0.18]{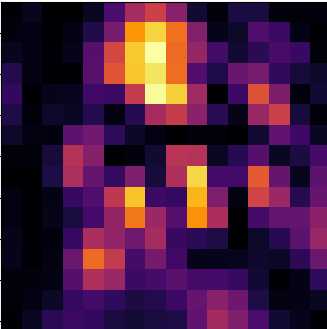}} node[left, pos=-0.02] {images$\,\,\,\,\,\,$};
                \draw[line,thick] (P |- environment.170) -- ++ (-1.2,0) |- (agent.190)
                node[right,pos=0.25,align=left] {reward\\ $r_t = -(1 - \mathnormal{SR}(o_{t}))^{2/5}$};
            \end{tikzpicture}
            \caption{Diagram of the interaction between the agent and its environment. At each time step $t$, the environment emits the observation $o_t$, containing the in-focus (top) and out-of-focus (bottom) images, as well as the reward $r_t$. The agent responds with action $a_t$ (the DM command) which prompts the environment to emit a new observation and reward $o_{t+1}$ and $r_{t+1}$, and so on. This figure is a reproduction of the one presented in Gutierrez et al. (2024\cite{Gutierrez-a-24}).}
            \label{fig:rlwfc}
        \end{figure}
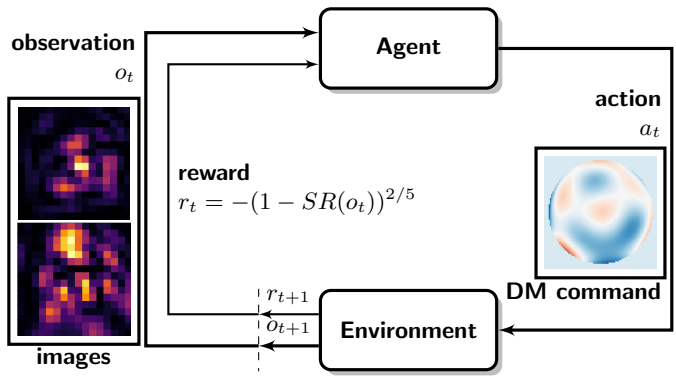

        In our formulation of the focal plane wavefront correction problem as an episodic MDP, illustrated in Figure \ref{fig:rlwfc}, the agent acts as the controller for the DM. The environment encompasses everything outside this controller, including the instrument dynamics and the incoming distorted wavefront. The state $s_t$ corresponds to the pupil plane phase $\phi$, which cannot be directly measured. Instead, the agent observes $o_t$ in the form of two focal plane phase diversity images, which contain all necessary information to estimate $\phi$. The agent uses this observation $o_t$ to select an action $a_t$, where in this context, the action represents the phase applied to the surface of the DM, $\phi_\text{DM}$. The reward $r_{t+1}$ is computed based on the subsequent observation and serves as a measure of the correction quality. Specifically, we employ a function of the Strehl ratio $\mathnormal{SR}(o_{t+1})$, which is a measure of the quality of the in-focus image, to compute the reward:
        
        \begin{equation}
            r_{t+1} = -(1 - \mathnormal{SR}(o_{t+1}))^{2/5} 
        \end{equation}
        
        This reward function design encourages the agent to minimize aberrations rapidly since a non-zero reward at each time step acts as a penalty, reducing the cumulative score the agent aims to maximize. The shape of the function incentivizes exploration when the Strehl ratio is low and tends towards more exploitative actions as the Strehl ratio approaches 1 (the maximum achievable value, corresponding to the absence of aberrations). This particular reward function was the best-performing out of the many that were trialed.
        
        Regarding the discount factor $\gamma$, which balances how much the agent values future rewards compared to immediate ones, we experimented with various values and found $\gamma=0$ to be optimal for training performance. This choice is consistent with the expected behavior of phase diversity, since each observation (consisting of 2 images, in focus and defocused) provides enough information to completely correct the aberrations with a single DM command, meaning we are merely interested in the Strehl ratio immediately following each action. If observations were limited to a single focal plane image, a higher $\gamma$ closer to 1 would likely be preferable. This adjustment would compel the agent to learn to create its own diversity image over time, prioritizing improved aberration estimates despite initial sub-optimal actions.
        
    \subsection{Training algorithm}
    
        The agent in this study is trained using the Proximal Policy Optimization (PPO) algorithm \cite{SchulmanWDRK17}, a variant of actor-critic \cite{konda1999actor}. The rationale behind the choice of algorithm is detailed in Section 3.2 of Gutierrez et al. (2024\cite{Gutierrez-a-24}). 
        
        Training is performed using the PPO implementation from the \texttt{Stable-baselines3} library \cite{stable-baselines3} in Python, for a fixed amount of time steps depending on the experiment. The parameters are optimized using the Adam algorithm \cite{kingma2017adam}, with an initial learning rate of $10^{-3}$ which is linearly decayed over the course of training.
        
        The chosen hyperparameters for PPO include a rollout buffer size of 8000 time steps, and a mini-batch size of 500 time steps. The default values from \texttt{Stable-baselines3} were kept for the remaining parameters (clipping parameter, epochs, Generalized Advantage Estimator decay parameter, advantage normalization, value function coefficient and entropy coefficient), as they demonstrated good performance in expermiental trials.
        
        Both the actor (the agent's  parameterized policy) and critic (the component that learns the value of actions in order to ``criticize'' the agent's choices) are implemented as neural networks (see reference~\citeonline{Gutierrez-a-24} for more details), sharing identical architectures with differing output layers. Inputs consist of flattened phase diversity images concatenated with the previous DM command. Hidden layers employ hyperbolic tangent activation functions. The output layer is designed to match the number of DM actuators for the actor's output layer, while the critic's output layer contains a single neuron.

\section{Numerical experiment in the non-coronagraphic case}
\label{sec:noncoro}

    In this section, we briefly describe the reference experiment from Gutierrez et al. (2024\cite{Gutierrez-a-24}), and showcase the agent's learning process on this experiment.

    \subsection{Parameters of the numerical experiment}

        The environment is implemented using OpenAI's Gym package \cite{brockman2016openai}, integrated with simulations managed through the Python-based Asterix package \cite{mazoyer_AsterixSimulator}. Observations are conducted at a wavelength of $\lambda=500\,\text{nm}$.
           
        The telescope features a circular, unobstructed pupil, imaging on a $16 \times 16$-pixel grid with a pixel pitch set to achieve Nyquist-Shannon sampling. The detector field of view is $8 \times 8$ $\lambda/D$.
        
\begin{figure}[htbp!]
            \centering\includegraphics[width=\linewidth]{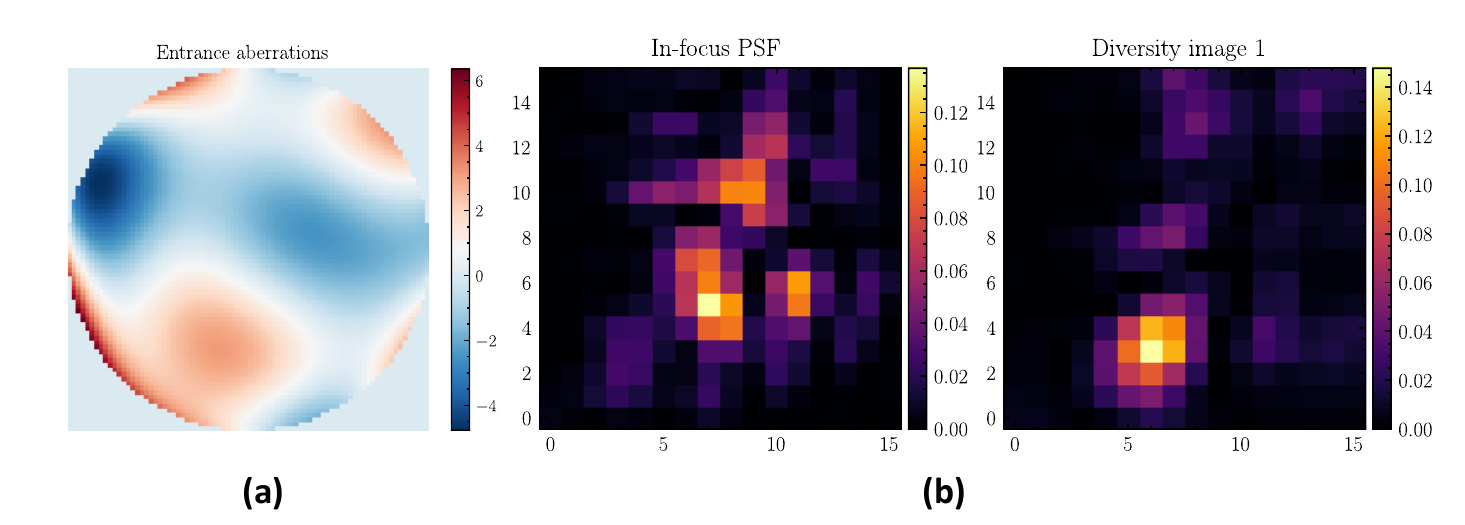}
            \vspace*{-.5\baselineskip}
            \caption{Example of an entrance phase map (a) and the resulting in-focus (left) and out-of-focus (right) images observed by the agent (b). Adapted from Gutierrez et al. (2024\cite{Gutierrez-a-24}).}
            \label{fig:examples}
        \end{figure}

        In this experiment, the entrance aberrations contain 21 Zernike modes, following a $1/f^2$ power spectral density (PSD) profile, which is typically encountered in good quality optics \cite{dohlen2011}. The total RMS wavefront error is set to 125 nm ($\lambda/4$) before correction, with tip/tilt limited to $+/- 0.5$ $\lambda/D$. Each training episode uses a new, independent phase screen (example in Fig.~\ref{fig:examples}(a)).

        For this experiment, each episode spans 4 time steps. The entrance aberrations are considered static for the duration of an episode. This choice, as well as the influence of episode length on learning efficiency, are discussed extensively in section 5.2 of Gutierrez et al. (2024\cite{Gutierrez-a-24}).
        
        A defocus aberration with a root mean square (RMS) amplitude of $\lambda/(2\sqrt{3})$ is introduced in the pupil plane to generate out-of-focus images. All images are normalized with respect to maximum intensity under ideal conditions. Fig.~\ref{fig:examples} (b) shows the diversity images obtained from the phase map in Fig.~\ref{fig:examples} (a).

        Correction is executed by a single ideal Zernike corrector DM with a number of actuators matching the number of modes in the aberrations (i.e., 21 in this case), capable of fully counteracting entrance aberrations when commanded correctly. This hypothetical DM has an instantaneous response time.
        
        We evaluate our method's noise tolerance by simulating a photon noise-limited detection system with an additional Gaussian readout noise with a standard deviation of one electron. We define a global signal-to-noise ratio, which we denote as SNR for brevity, as the signal-to-noise ratio in the brightest pixel in the un-aberrated case. The SNR is set to 100 in the reference experiment.
        
        This structured experiment framework allows us to rigorously assess our method's effectiveness in a controlled setting. This facilitates pinpointing training issues and optimizing performance.

    \subsection{Main result}
    
        The agent was trained for 10 million time steps (2.5 million episodes) on a 56-core CPU with a clock speed of 2.4 GHz and an Nvidia Tesla P100-PCIE-12GB GPU, taking about 10 hours.
        
\begin{figure}[htbp!]
            \centering\includegraphics[width=\myfraction\linewidth]{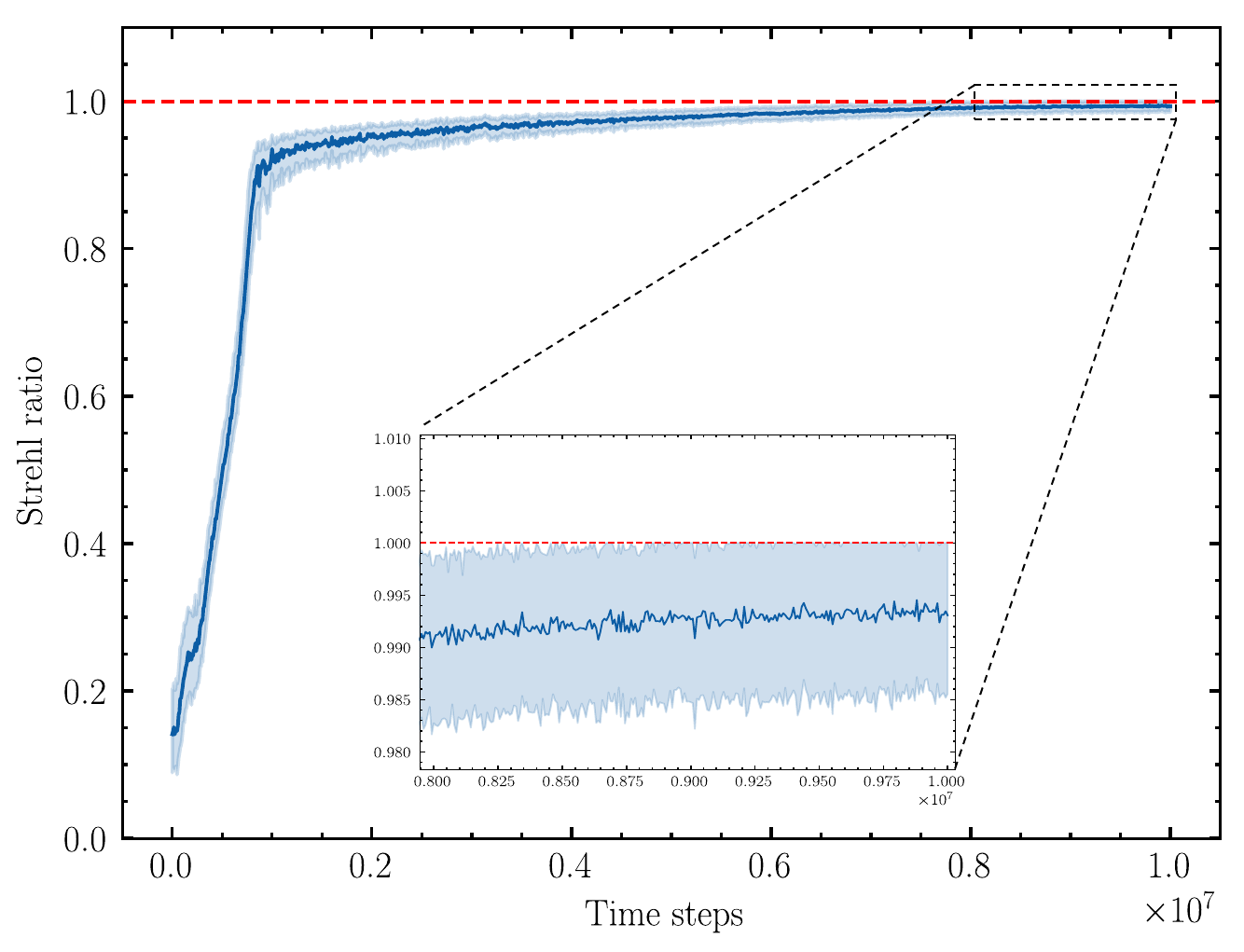}
            \vspace*{-.5\baselineskip}
            \caption{Average Strehl ratio after correction on the benchmark as a function of the number of training time steps. The error bands show the spread of the results between the 25th and 75th percentiles. The agent's performance gets progressively closer to the optimal theoretical performance (represented by the red dashed line), reaching an average Strehl ratio greater than 0.99 after 10 million time steps. This figure is taken from Gutierrez et al. (2024\cite{Gutierrez-a-24}).}
            \label{fig:learningcurve}
        \end{figure}

        Figure~\ref{fig:learningcurve} shows the agent's learning curve, which represents the evolution of its performance as training progresses (averaged over 5 runs). The agent is evaluated on a fixed benchmark set of 100 phase screens throughout training. The agent learns to achieve an average Strehl ratio of 0.9 after 2 million training time steps, ultimately exceeding 0.99 as it approaches 10 million training time steps. The linear decay applied to the global learning rate prevents overfitting and catastrophic forgetting (performance collapse). 
        
        This demonstrates our method's ability to autonomously learn aberration correction (21 modes) under realistic conditions (SNR=100, RMS wavefront deviation of $\lambda/4$) using only observations, without relying on prior knowledge about the aberrations.

        More numerical experiments and results can be found in Gutierrez et al. (2024\cite{Gutierrez-a-24}).

        Encouraged by these results, demonstrating the feasibility of a data-driven RL approach for non-coronagraphic wavefront control, we apply our method to a coronagraphic environment in the next section.

\section{Preliminary results of dark hole creation with reinforcement learning in simulations}
\label{sec:coro}

    This section details the adaptations required for our method to handle post-coronagraphic wavefront control. Many methods have been developed to estimate the electrical field creating the speckles after a coronagraph and correct for it using a deformable mirror \cite{GalicheretMazoyerreview}. The specificity of our approach is that we handle both estimation and correction simultaneously in a model-free approach. We present the initial numerical experiment designed for testing and showcase preliminary results.
    
    \subsection{Parameters of the numerical experiment}
        \subsubsection{Optical setup}
        
            In order to explore the applicability of our method to high-contrast imaging, we incorporate a coronagraphic system into our simulated telescope. This involves adding a focal plane equipped with a focal plane mask, which here is a four quadrant phase mask\cite{Rouan_2000}. Downstream from the focal plane mask, a pupil plane is introduced containing a Lyot stop with a diameter 97\% that of the entrance pupil. In the event of no aberrations, this combination of focal plane mask and Lyot stop prevents all of the on-axis light from reaching the imaging focal plane \cite{abe2003phase}.

\begin{figure}[htbp!]
                \centering\includegraphics[width=\linewidth]{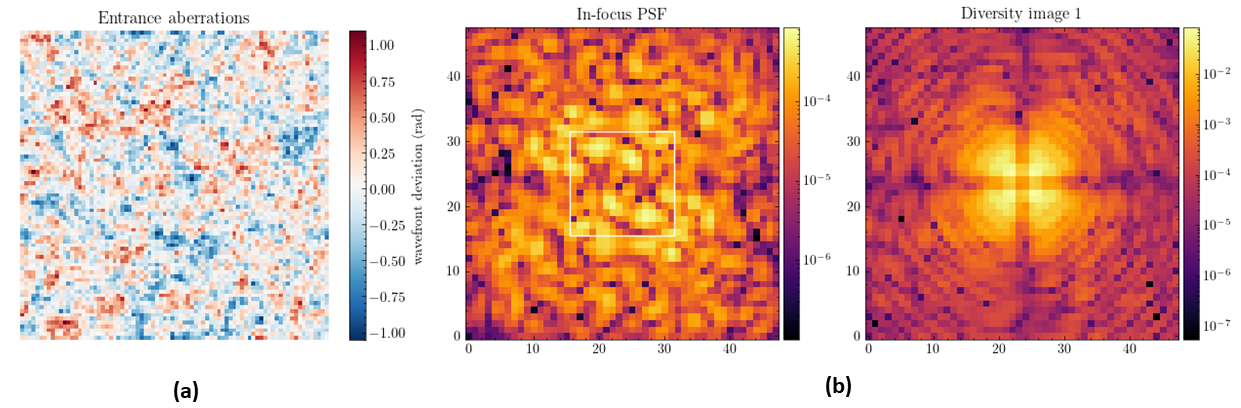}
                \vspace*{-.5\baselineskip}
                \caption{Example of an entrance phase map with the new representation (a) and the resulting in-focus (left) and out-of-focus (right) coronagraphic images observed by the agent (b). The white square outlines the area controlled by the DM (the dark hole).}
                \label{fig:coro_example}
            \end{figure}

            Since the Zernike basis is not practical for modelling very high order aberrations, which are commonplace in coronagraphic systems, we now create phase screens randomly (while still following the $1/f^2$ PSD profile) on a pixel basis in the pupil plane. Figure \ref{fig:coro_example} shows an example of such a phase map, and the resulting observation. To mimic exoplanet observations from a space telescope, the wavefront error is scaled down to 25 nm RMS  ($\lambda / 20$). The aberrations are again considered static for the duration of an episode, which now lasts for 10 time steps.
          
            The detector is enlarged to $48 \times 48$ pixels. The Zernike DM is replaced by a Fourier corrector, capable of controlling 64 Fourier modes in the pupil plane, enabling it to modulate the intensity of individual pixels in a dark hole of $8 \times 8$ $\lambda/D$ in the focal plane. The maximum peak-to-valley amplitude of the sine shapes applied to the surface of the DM is 800 nm for each mode. The defocus used to generate the diversity image is left unchanged from the non-coronagraphic case. However, it could benefit from being scaled down in the same way as the entrance aberrations\cite{Lee:99}. 

            To simplify the analysis in this early stage of development, we assumed ideal conditions with no noise sources, meaning that $SNR = \infty$.

            All other environmental parameters, such as the observation wavelength, were maintained at their non-coronagraphic values.
            
        \subsubsection{Training parameters}

            Because of the presence of the coronagraph, we need another metric than the Strehl ratio for evaluating the performance of the agent, as well as another reward. 
            
            The performance metric becomes the average normalized intensity in the dark hole, i.e. the average of the pixel intensities divided by the intensity in the brightest pixel in the un-aberrated non-coronagraphic point spread function (PSF). 
            
            To incentivize the agent to minimize the intensity in the dark hole, the reward $r_t$ at time step $t$ becomes:

            \begin{equation}
                r_t = -\log\left(\frac{\sum_{(x,y) \in DH} I_t(x,y)}{\sum_{(x,y) \in \mathcal{D}} PSF(x,y)}\right)/10 \ ,
            \end{equation}

            where $DH$ is the area inside the dark hole, $\mathcal{D}$ is total area of the detector, $I_t$ is the intensity recorded in the pixel of coordinates $(x,y)$ at time step $t$, and $PSF$ is the intensity of the un-aberrated PSF. While this function has yielded the best results so far, we are actively exploring alternative reward functions. This current implementation is likely to undergo significant modifications as we evaluate more promising candidates.

            The factor $1/10$ is there so that the reward stays below 1 as long as the averaged normalized intensity in the dark hole is over $10^{-10}$.

            For simplicity, the other training parameters, including the optimizer, learning rate, actor and critic architectures, and PPO hyperparameters, are left unchanged from the non-coronagraphic case. However, although this combination of parameters proved effective in the non-coronagraphic case, this may not represent the optimal configuration for the coronagraphic scenario. We plan to explore the parameter space further and potentially replace PPO with a more appropriate RL algorithm if a superior alternative exists.
        
    \subsection{Results}

\begin{figure}[htbp!]
            \centering\includegraphics[width=\myfraction\linewidth]{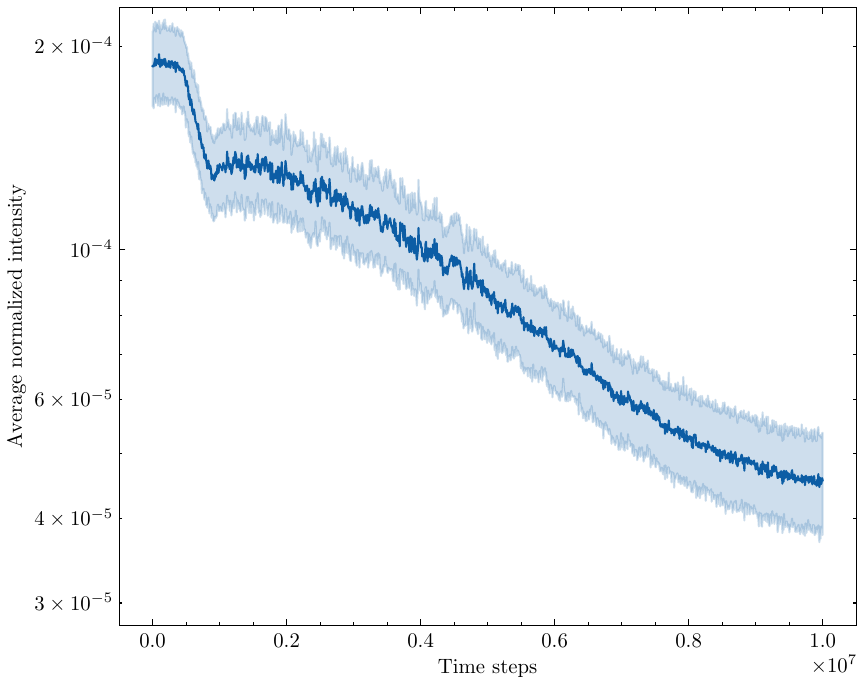}
            \vspace*{-.5\baselineskip}
            \caption{Average normalized intensity in the dark hole as a function of training time steps. The error bands represent the spread of the benchmark evaluation data between the 25th and 75th percentiles.}
            \label{fig:coro_learning_curve}
        \end{figure}

        Figure \ref{fig:coro_learning_curve} depicts the average learning curve of the agent across five training runs, with each run lasting 10 million time steps. The curve demonstrates the agent's ability to learn dark-hole digging. Over the training period, the average normalized intensity in the dark hole after correction is reduced from an initial value of $2 \times 10^{-4}$ to $4 \times 10^{-5}$. However, the curve also suggests that the agent's performance hasn't stopped improving after 10 million training steps. This indicates the potential for further improvement in two directions: on the one hand, a deeper dark hole is achievable with extended training; on the other hand, since we required much fewer training time steps to converge in the non-coronagraphic case (see Section~\ref{sec:noncoro}), we can hope that a better tuning of the training parameters would allow a faster convergence.

\begin{figure}[htbp!]
            \centering\includegraphics[width=\myfraction\linewidth]{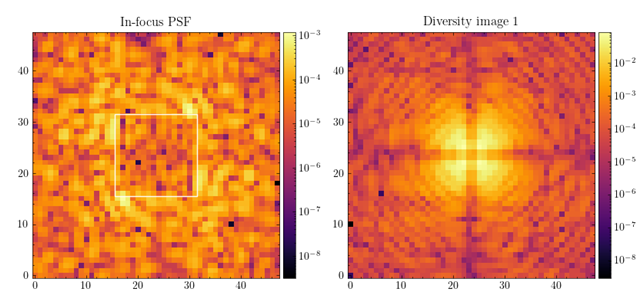}
            \vspace*{-.5\baselineskip}
            \caption{Example of observation at the end of an episode, after 10 correction time steps from the trained agent.}
            \label{fig:correction_example}
        \end{figure}

        Figure \ref{fig:correction_example} displays the diversity images obtained at the end of an episode, for the initial entrance aberrations presented in Figure \ref{fig:coro_example}.

        While these initial results demonstrate the potential of our fully data-driven approach, further development is necessary to achieve performance comparable to state-of-the-art wavefront control methods. This work represents a promising first step, as the agent successfully learns to create a dark hole without any external guidance. However, it's important to acknowledge the preliminary nature of this work, as development is in its early stages. We anticipate significant improvements through optimization of the reward signal, the training algorithm and its hyperparameters, and the neural network architectures.

\section{Conclusion and perspectives}

    Direct imaging of exoplanets necessitates the suppression of starlight using coronagraphs. However, aberrations on the optical path make residual starlight leak through the coronagraph and reach the imaging focal plane, significantly degrading the quality of scientific images and hindering exoplanet detection.  While existing aberration correction methods improve image quality, their iterative nature may lead to high execution times. Additionally, reliance on physical models introduces potential errors, limiting correction precision. To address these challenges, we developed a model-free RL approach, leveraging phase diversity images to directly control a DM, in the hope of reducing the number of image acquisitions required compared to traditional methods.

    We initially trained our RL agent on a simulated non-coronagraphic imaging scenario replicating realistic observational conditions. The agent achieved a Strehl ratio exceeding 0.99 with only 4 correction steps, after encountering 2.5 million independent phase screens during its training.

    Building upon the success achieved in the non-coronagraphic scenario, we transitioned to a simplified post-coronagraphic dark-hole digging task. Early results demonstrate the agent's capability to learn dark-hole creation. Preliminary results show that the agent is able to lower the average normalized intensity in the dark hole from $2 \times 10^{-4}$ to $4 \times 10^{-5}$, this time after 10 correction steps. These early yet promising results demonstrate the potential of our model-free approach to learn dark-hole creation. In this early stage of development, training parameters have yet to be optimized. These parameters were largely carried over from the non-coronagraphic case. Consequently, significant improvement is anticipated through future optimization efforts.

    Future work will focus on improving the depth of the dark-hole and reducing the number of correction iterations through careful optimization of the reward signal, as well as the neural network architectures, the training algorithm and its hyperparameters. Additionally, we will investigate more efficient diversity phases for generating diversity images. Following successful results in simulations, the next step will be experimental validation on the THD2 optical testbench \cite{baudoz2018statusperformancethd2bench}. Ultimately, our goal is to demonstrate the effectiveness of this method through on-sky observations.

\bibliographystyle{spiebib.bst}
\bibliography{yann,Acronymes,EnglishAcronyms}

\end{document}